\documentclass{article}

\usepackage{amssymb,amsfonts,amsmath,stmaryrd}
\usepackage{cite,enumerate,float,indentfirst}
\usepackage{color}

\def\be{\begin{eqnarray}}
\def\ee{\end{eqnarray}}
\def\nn{\nonumber}

\def\tr{{\rm tr}\,}
\def\Tr{{\rm Tr}\,}

\def\OP{\hbox{OP}}
\def\SP{\hbox{SP}}

\newcommand{\beq}{\begin{equation}}
\newcommand{\eeq}{\end{equation}}
\newcommand{\beqa}{\begin{eqnarray}}
\newcommand{\eeqa}{\end{eqnarray}}

\definecolor{red}{rgb}{1,0,0}
\definecolor{orange}{rgb}{1,0.5,0}
\definecolor{violet}{rgb}{0.7,0,1}

\hoffset= - 1.0in         

\begin{document}
\title{\bf
Superintegrability of
Kontsevich matrix model
}

\author{
{\bf Andrei Mironov$^{a,b,c}$}\footnote{mironov@lpi.ru; mironov@itep.ru},
\ and  \  {\bf Alexei Morozov$^{d,b,c}$}\thanks{morozov@itep.ru}
}
\date{ }

\maketitle

\vspace{-5.cm}

\begin{center}
\hfill FIAN/TD-08/20\\
\hfill IITP/TH-18/20\\
\hfill ITEP/TH-27/20\\
\hfill MIPT/TH-16/20
\end{center}

\vspace{3cm}

\begin{center}
$^a$ {\small {\it Lebedev Physics Institute, Moscow 119991, Russia}}\\
$^b$ {\small {\it ITEP, Moscow 117218, Russia}}\\
$^c$ {\small {\it Institute for Information Transmission Problems, Moscow 127994, Russia}}\\
$^d$ {\small {\it MIPT, Dolgoprudny, 141701, Russia}}
\end{center}

\vspace{.0cm}

\begin{abstract}
Many eigenvalue matrix models possess a peculiar basis of observables that have explicitly calculable averages. This explicit calculability is a stronger feature than ordinary integrability, just like the cases of quadratic and Coulomb potentials are distinguished among other central potentials, and we call it {\it superintegrability}. As a peculiarity of matrix models, the relevant basis is formed by the Schur polynomials (characters) and their generalizations, and superintegrability looks like a property $<character>\,\sim character$. This is already known to happen in the most important cases of Hermitian, unitary, and complex matrix models. Here we add two more examples of principal importance, where the model depends on {\it external fields}: a special version of complex model and the cubic Kontsevich model. In the former case, straightforward is a generalization to the complex tensor model. In the latter case, the relevant characters are the celebrated $Q$ Schur functions appearing in the description of spin Hurwitz numbers and other related contexts.
\end{abstract}

\section{Introduction}

In the original definition, matrix models are defined as averages over matrix
ensembles, often described by integrals over matrices or eigenvalues.
Generating functions of all correlators are then described by integrals
with arbitrary potentials in the action (matrix models of the first kind) or, alternatively, with
background fields (matrix models of the second kind).
In the both cases, one can prove that the partition functions are actually
$\tau$ functions of KP/Toda integrable theories, see \cite{MMint,versus,UFN3} for details,
we do not need them in the present text.

However, besides being just a $\tau$-function, the partition function of matrix model can be often presented as an explicit combinatorial power sum \cite{MM1,MM2}. This property is much similar to superintegrability of mechanical systems \cite{superint}, hence the term.

The superintegrability has been explicitly formulated so far for the matrix models of the first kind,
and, in this paper, our goal is to extend formulation to the second kind.
So far the known examples include:

\begin{itemize}
\item {\bf Rectangular complex model}

Correlators in the rectangular complex model are expressed \cite{AMMN,IMM,MM1,MM2} through the Schur polynomials (see earlier results in \cite{KPSS,Ramg}), which are characters of linear groups:
\be\label{Ccor}
\boxed{
\left<\chi_R\{\Tr (X\bar X)^k\}\right>:=
\int \chi_R\{\Tr (X\bar X)^k\} \cdot\exp\left(-\Tr X\bar X\right) d^2X \ =\
{  \chi_R\{N_1\}\ \chi_R\{N_2\}\over\chi_R\{\delta_{k,1}\}}
}
\ee
The matrix $X$ here is $N_1\times N_2$ rectangular matrix, $\chi_R\{p_k\}$ is the Schur polynomial (which is defined to be a symmetric function $\chi_R(z_i)$ of the  variables $z_i$, or a polynomial $\chi_R\{p_k\}$
of the power sums $p_k:=\sum_iz_i^k$), and the formula explains what is special about the Schur polynomials.
At the l.h.s. (\ref{Ccor}) the role of $z_i$ is played by the eigenvalues of the matrix $X\bar X$,
while at the r.h.s. $p_k=N_1$, $N_2$ or $\delta_{k,1}$.
Averages of $\tr (X\bar X)^k$ {\it per se}
are more involved than those of characters
and actually contain an additional summation over Young diagrams at the level $|R|$,
for examples of the corresponding Harer-Zagier formulas, see \cite{HZ}.
In (\ref{Ccor}) and below, all the integration measures are normalized to unity, $<1>\,:=1$;
in other words, we omit normalization integrals in denominators to simplify formulas.

For the quantity $\chi_R\{\delta_{k,1}\}$, there is a hook formula \cite{Fulton}
\be\label{hook}
\chi_R\{\delta_{k,1}\}=\frac{1}{\prod_{(\alpha,\beta)\in R} h_{\alpha,\beta}}
= {\prod_{i<j}^{l_R} (R_i-i-R_j+j)\over \prod_i^{l_R} (l_R+R_i-i)!}
\ee
where  $l_R$ is the number of lines $R_i$ in the Young diagram $R$: $R_1\ge R_2\ge\ldots\ge R_{l_R}$, and $|R|$ is the size of the Young diagram $R$: $|R|:=\sum_{i=1}^{l_R}R_i$,
and $h_{\alpha,\beta}$ is the hook length for a box  $(\alpha,\beta)\in R$.
For other similar loci  we have,
up to a coefficient, equal to $\pm 1$ and $0$,
\be
\chi_R\{\delta_{k,s}\}\sim \prod_{(\alpha,\beta)}\frac{1}{[[h_{\alpha,\beta}]]_{s,0}}
\ee
where we use the notation
\be
[[n]]_{s,a} = n\ {\rm if}\  n=a \,{\rm mod}(s) \ \  {\rm and}  \ 1 \ {\rm otherwise}
\label{sa}
\ee
which will be also useful beyond Gaussian models.
Also simple in terms of the box coordinates is the expression   on the topological locus:
\be
\chi_R\{N\}= \chi_R\{\delta_{k,1}\}\prod_{(\alpha,\beta)\in R}(N+\alpha-\beta)
\label{toploc}
\ee

Using eq.(\ref{Ccor}) and the Cauchy formula
\be\label{Cauchy}
\exp\left(\sum_k {p_k\over k}\ \Tr (X\bar X)^k\right) =\sum_R \chi_R\{p_k\}\cdot\chi_R\{\Tr (X\bar X)^k\}
\ee
one can immediately obtain a combinatorial expression for the generating function of all correlators:
\be\label{CP}
\int \exp\left(-\Tr X\bar X+\sum_k {p_k\,\over k}\Tr (X\bar X)^k\right) d^2X
=\ \sum_R{ \chi_R\{p_k\}\ \chi_R\{N_1\}\ \chi_R\{N_2\}\over\chi_R\{\delta_{k,1}\}}
\ee

\item {\bf Gaussian Hermitian model}

Correlators in the Gaussian Hermitian model are \cite{IMM,MM1,MM2} (cf. also with a Fourier expansion of \cite[Eq.(2.18)]{MKR} in terms of characters)
\be\label{Gcor}
\boxed{
\left<\chi_R\{\Tr X^k\}\right>:=
\int \chi_R\{\Tr X^k\} \cdot\exp\left(-{1\over 2}\Tr X^2\right) dX =
{  \chi_R\{N\}\ \chi_R\{\delta_{k,2}\}\over\chi_R\{\delta_{k,1}\}}
}
\ee
where $X$ is an $N\times N$ matrix, and the generating function of all correlators is
\be\label{GP}
\int \exp\left(-{1\over 2}\Tr X^2+\sum_k{p_k\over k}\ \Tr X^k\right) dX =
\sum_R{ \chi\{p_k\} \chi_R\{N\}\ \chi_R\{\delta_{k,2}\}\over\chi_R\{\delta_{k,1}\}}
\ee

\item {\bf Trigonometric (unitary type) matrix model}

This model is given by the eigenvalue integral, and the correlators are \cite{MMsum}
\be
\boxed{
\left<\chi_R(e^{m_i})\right>:=
\int \chi_R(e^{m_i}) \prod_{i<j}^N \sinh^2\! \left(\frac{m_i-m_j}{2}\right)
\prod_{i=1}^N \exp\left({-\frac{m_i^2}{2g^2}}\right) dm_i \
= \  A^{|R|}\cdot q^{2\varkappa_R } \cdot \chi_R\{p^*\}
}
\ee
When averaging, the argument of character in the integrand is the diagonal matrix
with the entries $e^{m_i}$,
and, at the r.h.s., the parameters are
$q = e^{g^2/2}$, $A=q^N=e^{Ng^2/2}$, the exponent $\varkappa_R = \sum_{(x,y)\in R}(y-x)$, and the time variables $p^*_k = \frac{A^{k}-A^{-k}}{q^k-q^{-k}}$ in the argument
of the character at the r.h.s. lie in the
``topological locus'' obtained by the $q$-deformation of $p_k = N$ in (\ref{Gcor}).

\item {\bf Knot matrix model}

Further deformation of the measure to the knot eigenvalue model gives basically the same expression for the correlators \cite{BEM}:
\be
\boxed{
\left<\chi_R(e^{m_i\over a})\right>:=
\int \chi_R(e^{m_i\over a}) \prod_{i<j} \sinh \frac{m_i-m_j}{2a} \sinh\frac{m_i-m_j}{2b}
\prod_{i=1}^N e^{{-\frac{m_i^2}{2g^2}}}\, dm_i
=   \left(A^{|R|}\cdot q^{2\varkappa_R }\right)^{b/a}\!\! \cdot \chi_R\{p^*\}
}
\ee
with $q=e^{\frac{g^2}{2ab}}$.

The correlators of $\left< \chi_R(e^{m_i}) \right>$ are more involved this time, but are related to the HOMFLY-PT polynomials of the torus knots ${\cal H}_R^{{\rm Torus}_{a,b}}(A,q)$ \cite{BEM}:
\be\label{52}
\left< \chi_R(e^{m_i}) \right>
 = {\cal H}_R^{{\rm Torus}_{a,b}}(A,q)
 = \chi_R\{N\} \cdot H_R^{{\rm Torus}_{a,b}}(A,q)
\ee
where
\be
{\cal H}_R^{{\rm Torus}_{a,b}}(A,q) =
\left< \sum_{Q\vdash a|R|} c_{R,Q} \chi_Q(e^{m_i\over a}) \right>
= A^{\frac{b|R|}{a}}\sum_{Q\vdash a|R|} c_{R,Q}
\cdot q^{\frac{2b \varkappa_Q}{a} } \cdot \chi_Q\{p^*\}
\ee
where $c_{R,Q}$ are peculiar Adams coefficients \cite{RJ}.

Generalization to non-torus knots, for example, like that in \cite{AlexMMM},
remains to be analyzed.

\item {\bf $q,t$ deformed Gaussian Hermitian model}

The eigenvalue $q,t$-deformed Gaussian Hermitian model is associated with a deformation of Schur polynomials:
all of them  in the formula (\ref{Gcor}) are replaced by the corresponding Macdonald polynomials $M_R$ \cite{MPS}:
\be
\boxed{
\left<M_R(z_i)\right>:=
\!\int_{-1}^1 \!\! M_R(z_i)\prod_{j\ne i}{\left({z_i\over z_j};q\right)_\infty\over
\left(t{z_i\over z_j};q\right)_\infty} \prod_i z_i^{(N-1)\log_qt}(q^2z_i^2;q^2)_\infty d_qz_i =
{M_R\Big\{{1-t^{Nk}\over 1-t^k}\Big\}\
M_R\Big\{{1+(-1)^k\over 1-t^k}\Big\}\over M_R\Big\{{1\over 1-t^k}\Big\}}
}
\ee
where
\be
(z;q)_\infty:=\prod_{k=0}^{\infty}(1-q^kz)
\ee
and the integral is defined to be the Jackson integral,
\be
\int_{-1}^1 d_qz f(z):=(1-q)\sum_{k=0}^\infty q^k\Big(f(\xi q^k)+f(-\xi q^k)\Big)
\ee
The measure $d\mu_{q,t}(z)=(q^2z^2;q^2)_\infty d_qz$ gives rise to the Gaussian measure $d\mu_G(z)=\exp\Big(-{1\over 2}z^2\Big)dz$ in the limit of $q\to 1$.

One can expect also a further extension of similar character identities to the elliptic $q,t$ model  \cite{MMell}.

\item{{\bf Monomial non-Gaussian models}}

Superintegrability is in no way restricted to the Gaussian integration measures or their deformations,
though  beyond them the knowledge is still restricted.
The most important example is provided by {\it monomial} non-Gaussian actions:
\be
\boxed{
\begin{array}{c}
\left<\chi_R\{\Tr X^k\}\right>_a:=
\int_{C_{s,a}^{\otimes N}} \chi_R\{\Tr X^k\} e^{-\Tr X^s/s} dX =\cr\cr
=\chi_R\{\delta_{k,s}\}\cdot
\prod_{(\alpha,\beta)\in R} [[N+\alpha-\beta]]_{s,0}\cdot [[N+\alpha-\beta]]_{s,a}\cr\cr
N=0\ \hbox{or}\ a\ \hbox{mod}\ s
\end{array}}
\label{monomodels}
\ee
where we used the notation from (\ref{sa}).
The new trick here is the use of a special star-like (closed) integration contour $C_{s,a}$:
\be
\int_{C_{s,a}}  F(x)\ e^{-x^s/s} \,dx \ :=\ \sum_{b=1}^s e^{-2\pi i(a-1)b/s}\cdot \int_0^\infty
F\big(e^{2\pi ib/s}x\big)\ e^{-x^s/s}\, dx
\ee
which picks up only the  powers of $x$, which are equal to $a-1\, {\rm mod}\, s$, in particular,
\be
\int_{C_{s,a}} x^ke^{-x^s/s} dx=\delta_{k+1-a}^{(s)}\cdot\Gamma\left({k+1\over s}\right)
\ee
$\delta_k^{(s)}$ is defined to be 1 if $k=0$ mod $s$ and to vanish otherwise.
This makes the answer depending on an additional parameter $a=0,\ldots s-1$. The r.h.s. of (\ref{monomodels}) contains some factors $N+j$ from $\chi_R\{N\}$, (\ref{toploc}), that is,
those with $N+i = 0,a\, {\rm mod}(s)$, and, hence, its vanishing depends also on the value of $N$.
Note that if the condition
$N=0\ \hbox{or}\ a\ \hbox{mod}\ s$ is not satisfied, one can not define the correlator by the condition $<1>=1$ because of zeroes in the denominator.
Note also that the Vandermonde determinant in the integrand is independent of $s$ and $a$.
For more details, see the original paper \cite{PSh}.

\end{itemize}

\bigskip

In this paper, we extend these results to the simplest background field models:
the matrix integral over $N_1\times N_2$ complex matrices $X$ with the measure
\be
\mu_C(X) \sim e^{-\tr AXB\bar X} d^2X
\ee
where $N_1\times N_1$ matrix $A$ and $N_2\times N_2$ matrix $B$ are fixed external matrices,
and the Hermitian generalized Kontsevich model \cite{GKM}
in background matrix field $\Lambda$,
i.e. the matrix integral over $N\times N$ Hermitian matrix $X$ with the measure
\be
\mu_K(X) \sim e^{\tr(W(X) + \Lambda X)} dX
\ee
and $\Lambda$ is an external $N\times N$ matrix. In this paper, we consider these two examples, the potential in the second one being $W(X) = -\frac{X^3}{3}$, which is just the original Kontsevich model \cite{Kon}.
Remarkably, the difference between these two models is severely increased
in this case:
in particular, in the Hermitian case, only one background field can be easily handled.

The question we address is what are the functions of $X$-variables
that have simple and explicitly calculable averages.
As we demonstrate in this paper, in the complex case, these are still the Schur polynomials $\chi_R\{\tr (X\bar X)^k\}$,
however, the r.h.s. of (\ref{Ccor}) now contain traces of $A$ and $B$ instead of $N_1$ and $N_2$:
\be
\boxed{
\left<\chi_R\{\Tr (X\bar X)^k\}\right> =
\frac{\chi_R\{\tr A^{-k}\} \chi_R\{\tr B^{-k}\}}{\chi_R\{\delta_{k,1}\}}
}
\label{compAB}
\ee
see sec.\ref{RCM} below for details, and for generalization to the complex {\it tensor} model.
In the Hermitian (Kontsevich) case, the story makes a new twist:
the relevant functions are more restricted $Q$ Schur polynomials,
see sec.\ref{Cccor} for the correlators in this case,
\be
\boxed{
\left<Q_R\{\Tr X^k\}\right> \ =\
\delta_{R,even}\cdot
{Q_{R/2}\{\Tr\Lambda^{-k}\}Q_{R/2}\{\delta_{k,1}\}\over Q_{R}\{\delta_{k,1}\}}
}
\ee
and sec.\ref{Kon} for the combinatorial expression for the Kontsevich model
\be
\boxed{\boxed{
Z_K=\!\!
\sum_{R\in \SP}{1\over 4^{|R|}}
{Q_{R}\{\Tr\Lambda^{-k}\}Q_{R}\{\delta_{k,1}\}Q_{2R}\{\delta_{k,3}\}\over Q_{2R}\{\delta_{k,1}\}}
}}
\label{main}
\ee
Contributing here are only diagrams $R$ with all lines of even length.
Eq.(\ref{main}) can be considered as the main new result of this paper.
Important feature of formulas (\ref{compAB})-(\ref{main}) is that they depend on the matrix size
only implicitly, through the traces of powers of background fields.
This is the crucial feature, which allows one to forget about the matrix-integral origin/realization
of these models, in particular to treat them in terms of $\tau$-functions of integrable hierarchies \cite{GKM}.

\section{Rectangular complex model\label{RCM}}

\subsection{Correlators in terms of permutations}

We start with calculating the correlators in the complex matrix model with the external matrix (see \cite{LSZ} for a square matrix example), and use the symmetry group technique worked out in \cite{MM2}. In this section, we closely follow that paper.

We consider the rectangular complex model
\be
Z_C:=\int d^2X\exp\left(-\Tr AXB\bar X \right)
\ee
with correlators defined as
\be
\left<{\cal O}(X)\right>:=
\int  {\cal O}(X)\exp\left(-\Tr AXB\bar X \right) d^2X
\ee
where $X$ is $N_1\times N_2$ rectangular matrix, $A$ and $B$ are square matrices of sizes $N_1\times N_1$ and $N_2\times N_2$ respectively, and (notice that, with this definition, the matrix multiplication is defined by convolution of the first indices with each other, and of the second indices with each other)
\be
\Tr AXB\bar X:=\sum_{ijk}A^i_kX_{ij}B^j_{l}\bar X^{kl}
\ee
In order to calculate the pair correlator is equal to
\be\label{pair}
\left<X_{ij}\bar X^{kl}\right>=\Big(A^{-1}\Big)_{ik}\Big(B^{-1}\Big)_{jl}
\ee
The $2m$-point correlator can be labelled by a permutation of indices $\sigma$ that belongs to the symmetric group $S_m$:
\be
\left<{\cal O}_\sigma\right> = \left< \prod_{p=1}^{l} \Tr (X\bar X)^{m_p} \right>
= \left<   \prod_{i=1}^m  X_{a_i\alpha_i} {\bar X}^{a_i\alpha_{\sigma(i)}}\right>
\ee
In order to calculate this correlator, we apply the Wick theorem and use formula (\ref{pair}):
\be
\left<\prod_{i=1}^m X_{a_i\alpha_i} \bar X^{b_i\beta_{i}}\right>
= \sum_{\gamma \in S_m} \prod_{i=1}^m
\Big(A^{-1}\Big)_{a_{ i}}^{b_{\gamma(i)}}
\Big(B^{-1}\Big)_{\alpha_{ i}}^{\beta_{\gamma(i)}}
\label{Wick}
\ee
and obtain
\be
\left<{\cal O}_\sigma\right> =\sum_{\gamma \in S_m} p^A_{\gamma} p^B_{\gamma\circ\sigma}
\label{OasasumMatrix}
\ee
where
\be\label{times}
p^A_k:=\Tr A^{-k},\ \ \ \ \ \ \ p^B_k:=\Tr B^{-k},\ \ \ \ \ \ \ p^{A,B}_{\gamma}:=\prod_{i=1}^{l_\gamma}p^{A,B}_{i(\gamma)}
\ee
and $l_\gamma$ is the number of cycles in the permutation $\gamma$,
with $i(\gamma)$ denoting the $i$-th cycle in the permutation $\gamma$.

Now we use the standard identity \cite{Fulton}
\be
p_{\gamma}=\sum_{R\vdash |{\gamma}|}\psi_R({\gamma})\chi_R\{p\}
\ee
Here the sum goes over all Young diagrams $R$ with $|{\gamma}|$ boxes, $\chi_R\{p\}$ is the Schur polynomial (the character of the linear group $GL(N)$), $p_{\gamma}=\prod_{i=1}^{l_\gamma}p_{\gamma_i}$ and $\psi_R({\gamma})$ is the character of the symmetric group $S_{|R|}$.
Now using the orthogonality relation
\be\label{OR}
\sum_\gamma \psi_R(\gamma)\psi_Q(\gamma\circ\sigma)=\sum_\gamma \psi_R(\gamma^{-1})\psi_Q(\gamma\circ\sigma) =
\frac{\psi_R(\sigma)}{d_R}\, \delta_{QR}
\ee
where $d_R:=\chi_R\{\delta_{k,1}\}$, we finally come to
\be
\left<{\cal O}_\sigma\right> = \sum_{R\vdash m} \frac{\chi_R\{p^A\}\chi_R\{p^A\}}{\chi_R\{\delta_{k,1}\}}\psi_R(\sigma)
\label{ORCM}
\ee
where the sizes of $R$ and $\sigma$ coincide.

\subsection{Complex model in background fields}

Let us now use the Frobenius formula \cite{Fulton}
\be\label{F}
\chi_R\{p_k\}={1\over |R|!}\sum_\gamma\psi_R(\gamma)p_\gamma
\ee
where $|\gamma|$ is the size of the Young diagram $R$, and the orthogonality relation (\ref{OR}) with the particular value
$\psi_R(id)=\psi_R([1^{|R|}])=|R|!d_R$. Then, one immediately obtains from (\ref{ORCM})
\be
\boxed{
\left<\chi_R\{\Tr (X\bar X)^k\}\right> =
\frac{\chi_R\{\tr A^{-k}\} \chi_R\{\tr B^{-k}\}}{\chi_R\{\delta_{k,1}\}}
}
\ee
This generalizes the  answer (\ref{Ccor}) to the case of $A\neq I$, $B\neq I$.

Similarly to (\ref{CP}), we also can obtain the combinatorial expression for the generating function
\be
\boxed{
\int \exp\left(-\Tr AXB\bar X+\sum_k {p_k\over k}\Tr (X\bar X)^k\right) d^2X \
=\ \sum_R{ \chi_R\{p_k\}\chi_R\{\tr A^{-k}\} \chi_R\{\tr B^{-k}\}\over\chi_R\{\delta_{k,1}\}}}
\ee

\subsection{Tensor model in background fields}

The results of this section for the complex matrix model can be straightforwardly extended to the Gaussian tensor model in background fields. To this end, we follow \cite{IMM3} and again apply the technique of \cite{MM2}.

The Gaussian tensor model is a model of complex $r$-tensors $X_{a^1,...a^r}$ where each runs trough its interval $a_i=1,\ldots,N_i$, in the background of $r$ external square matrices $A_{(i)}$ of the sizes $N_i$ with the Gaussian action
\be
S:=\sum_{a^1,b^1=1}^{N_1}\ldots\sum_{a^r,b^r=1}^{N_r} X_{a^1,...a^r}\bar X^{b^1,...b^r}
\prod_{i=1}^r\Big(A_{(i)}\Big)_{b_i}^{a_i}
\ee
Gauge invariant operators at the level $m$ in any (not obligatory Gaussian) tensor model are the tensorial counterparts of ``multi-trace" operators
\be\label{gio}
{\cal O}_{\sigma_1,\ldots,\sigma_r}
= \sum_{\vec a^1=1}^{N_1}\ldots\sum_{\vec a^r=1}^{N_r}
\left(\prod_{p=1}^m X_{a^1_p,...a^r_p}
\bar X^{a^1_{\sigma_1(p)},\ldots,a^r_{\sigma_r(p)}}\right)
\ee
labeled by the set of $m$ elements $\sigma_i$  of the permutation group $S_m$.
In fact, the labeling is reduced to a double coset $S_m\backslash S_m^{\otimes r}/ S_m$ \cite{R,IMM2}, but we do not need these details here.

There is also a distinguished set of operators called generalized characters in \cite{IMM3} that are defined as
\be
\chi_{R_1,\ldots, R_r}(X,\bar X)
= \frac{1}{n!}\!\!\!\!\!\!\!\!\!\sum_{\ \ \ \ \ \sigma_1,\ldots,\sigma_r\in S_n}\!\!\!\!\!\!\!\!\!
\psi_{R_1}(\sigma_1)\ldots\psi_{R_r}(\sigma_r)\cdot {\cal O}_{\sigma_1,\ldots,\sigma_r}
\label{hurchar}
\ee
These generalized characters do not form a full basis in the space of all gauge invariant operators, instead they form an over-complete basis in the space of all gauge invariant operators with non-vanishing Gaussian averages \cite{IMM3}.

The pair correlator is now
$
\left<X_{a^1_p,...a^r_p}\bar X^{b^1_p,...b^r_p}\right>=\left(A_{(1)}^{-1}\right)_{a_1}^{b_1}\cdots\left(A_{(r)}^{-1}\right)_{a_r}^{b_r}
$
and, in complete analogy with what we were doing in the complex matrix model case, we obtain the averages of these generalized characters
\be\boxed{
\Big< \chi_{R_1,\ldots, R_r} \Big> \ =
\ {C_{R_1,\ldots,R_r}\over d_{R_1}\cdot\ldots \cdot d_{R_r}} \cdot
\chi_{R_1}\left\{\Tr A_{(1)}^{-k}\right\}\cdot \ldots\cdot \chi_{R_r}\left\{\Tr A_{(r)}^{-k}\right\}}
\ee
where
\be
C_{R_1,\ldots,R_r}:=\sum_{\Delta\vdash n}{\prod_{i=1}^r \psi_{R_i}(\Delta)\over z_\Delta}
\ee
and $z_\Delta:=\prod_k k^{m_k}m_k!$ is the standard symmetric factor of the Young diagram (order of the automorphism).
In the case of $r=3$, $C_{R_1,R_2,R_3}$ are the Clebsch-Gordan coefficients of the three irreducible representations $R_1$, $R_2$, $R_3$ of the symmetric group.

\section{Correlators in Hermitian model}

\subsection{Examples of correlators}

Let us calculate the correlators in the Gaussian Hermitian matrix model with external fields

Since the matrices are square, we return to the standard convention of convolution of indices in the matrix multiplication.
One could expect that, similarly to coming from the rectangular complex model to the Gaussian Hermitian one, the results will not be too much different from those in sec.\ref{RCM}. It turns out, however, not to be the case, and both the calculations in the Hermitian case with the external field are much more tedious even in the case of one external matrix, and the results remarkably differ from those of sec.\ref{RCM}.
We consider here only the case of one external matrix, i.e. the correlators of the form
\be\label{Kc}
\left<F(X)\right>:=
\int F(X) \exp\left(-\Tr X^2\Lambda\right)  dX
\ee
Since, the partition function is an invariant function of the external matrix $\Lambda$, i.e. depends only on its traces, one suffices to choose $\Lambda$ diagonal. Similarly to (\ref{times}), we define $p_k:=\Tr\Lambda^{-k}$.

In this model, the pair correlator in terms of eigenvalues $\lambda_i$ of the (diagonal) matrix $\Lambda$ is equal to
\be
\left<X_{ij}X_{kl}\right>={2\delta_{il}\delta_{jk}\over \lambda_k+\lambda_l}
\ee
Here is the difference with the case of two external matrices: in that case one can not work in terms of eigenvalues, and the pair correlator is much more involved.

Now we consider examples of simple correlators in order to demonstrate the structure of answers.

First consider $<\Tr X^3\cdot\Tr X^3>=<X_{ij}X_{jk}X_{ki}X_{\alpha\beta}X_{\beta\gamma}X_{\gamma\alpha}>$. This average is equal to the sum of products of pairings: $<X_{ij}X_{jk}><X_{ki}X_{\alpha\beta}><X_{\beta\gamma}X_{\gamma\alpha}>$ plus all possible other pairings. Totally, there are 15 different possibilities: 3 terms of the form
\be
C_1=<X_{ij}X_{\alpha\beta}><X_{jk}X_{\beta\gamma}><X_{ki}X_{\gamma\alpha}>=
\sum_{\alpha}{1\over \lambda_\alpha^3}=p_3
\ee
three terms of the form
\be
C_2=<X_{ij}X_{\alpha\beta}><X_{jk}X_{\gamma\alpha}><X_{ki}X_{\beta\gamma}>=\sum_{\alpha,\beta,\gamma}
{8\over (\lambda_\alpha+\lambda_\beta)(\lambda_\beta+\lambda_\gamma)(\lambda_\gamma+\lambda_\alpha)}
\ee
and nine terms of the form
\be
C_3=<X_{ij}X_{jk}X_{ki}X_{\alpha\beta}X_{\beta\gamma}X_{\gamma\alpha}>=\sum_{\alpha,\beta,\gamma}
{4\over \lambda_\alpha (\lambda_\alpha+\lambda_\beta)(\lambda_\alpha+\lambda_\gamma)}
\ee
Note that $C_2+3C_3=p_1^3/2$. Thus, we finally obtain $<\Tr X^3\cdot\Tr X^3>=3p_3+12p_1^3$.

Consider examples of all correlators up to the level 6 (for the sake of brevity, we use the notation $\xi_k:=\sum_{i,j}{1\over (\lambda_i+\lambda_j)^k}$):
\be\label{excor}
<\Tr X^2>=2\xi_1\nn\\
\boxed{<\Tr X\cdot\Tr X>=p_1}\nn\\
\nn\\
\hline\nn\\
\nn\\
<\Tr X^4>=8\sum_{i,j,k}{1\over (\lambda_i+\lambda_j)(\lambda_i+\lambda_k)}+p_2\nn\\
\boxed{<\Tr X^3\cdot\Tr X>=3p_1^2}\nn\\
<\Tr X^2\cdot\Tr X^2>=4\xi_1^2+8\xi_2\nn\\
<\Tr X^2\cdot\Tr X\cdot\Tr X>=2p_1\xi_1
+2p_2\nn\\
\boxed{<\Tr X\cdot\Tr X\cdot\Tr X\cdot\Tr X>=3p_1^2}\nn\\
\nn\\
\hline\nn
\ee
\be
\nn\\
<\Tr X^6>=16\sum_{i,j,k,l}{1\over (\lambda_i+\lambda_j)(\lambda_i+\lambda_k)(\lambda_i+\lambda_l)}+
24\sum_{i,j,k,l}{1\over (\lambda_i+\lambda_j)(\lambda_i+\lambda_k)(\lambda_k+\lambda_l)}+p_3+6p_1p_2\nn\\
\boxed{<\Tr X^5\cdot\Tr X>=5p_3+10p_1^3}\nn\\
<\Tr X^4\cdot\Tr X^2>=8p_1\sum_{i,j,k}{1\over (\lambda_i+\lambda_j)(\lambda_i+\lambda_k)}+64\sum_{i,j,k}{1\over (\lambda_i+\lambda_j)^2(\lambda_i+\lambda_k)}+p_1p_2+p_3\nn\\
<\Tr X^4\cdot\Tr X\cdot\Tr X>=8p_1\sum_{i,j,k}{1\over (\lambda_i+\lambda_j)(\lambda_i+\lambda_k)}+8\sum_{i,j}{1\over
\lambda_i^2(\lambda_i+\lambda_j)}+9p_1p_2\nn\\
\boxed{<\Tr X^3\cdot\Tr X^3>=3p_3+12p_1^3}\nn\\
<\Tr X^3\cdot\Tr X^2\cdot\Tr X>=6p_1^2\xi_1+12p_1p_2\nn\\
\boxed{<\Tr X^3\cdot\Tr X\cdot \Tr X\cdot \Tr X>=3(2p_3+3p_1^3)}\nn\\
<\Tr X^2\cdot\Tr X^2\cdot\Tr X^2>=8\xi_1^3+48\xi_1\xi_2+64\xi_3\nn\\
<\Tr X^2\cdot\Tr X^2\cdot\Tr X\cdot\Tr X>=
4p_1\xi_1^2+8p_1\xi_2
+8p_2\xi_1+8p_3\nn\\
<\Tr X^2\cdot\Tr X\cdot\Tr X\cdot\Tr X\cdot\Tr X>=6p_1^2\xi_1+12p_1p_2\nn\\
\boxed{<\Tr X\cdot\Tr X\cdot\Tr X\cdot \Tr X\cdot\Tr X\cdot\Tr X>=15p_1^3}
\ee
and all odd level correlators evidently vanish.

These examples demonstrate that the correlators that involve traces of only odd degrees of the matrix $X$ (they are boxed in (\ref{excor})) are expressed only through the times (\ref{times}), moreover, through the odd times. We will return to other correlators elsewhere, and here we consider only this type of correlators. We will need a set of polynomials of odd times that gives a complete basis and forms a closed ring.

\subsection{$Q$ Schur polynomials}

Emergency of only the odd times immediately gives one a hint to use the $Q$ Schur polynomials instead of the standard Schur polynomials, as we did in the previous sections. These polynomials\footnote{More detailed review of these polynomials can be found, e.g., in \cite{MMN}.} depend only on {\it odd} time-variables
$p_{2k+1}$ and only on {\it strict} Young diagrams $R=\{r_1>r_2>\ldots >r_{l_R}>0\}\in \SP$. They were introduced by I. Schur \cite{Schur} in the study of projective representations of symmetric groups, and were later identified by I. Macdonald \cite{Mac} with the Hall-Littlewood polynomials $\hbox{HL}_R\{p\}:=\hbox{Mac}_R\{p\}\Big|_{q=0}$ at $t=-1$:
\be
Q_R:=
\left\{\begin{array}{cl}
2^{l_{_R}/2}\cdot \hbox{HL}_R(t=-1)&\hbox{for}\ R\in\hbox{SP}\\
&\\
0&\hbox{otherwise}
\end{array}
\right.
\ee
Macdonald's observations were that
$\hbox{HL}_R\{p\}$ for $R\in \SP$ depend only on odd time-variables $p_{2k+1}$, and that
$\hbox{HL}_R\{p\}$ for $R\in \SP$ form a sub-ring, i.e. the Littlewood-Richardson coefficients ${\cal N}_{R_1,R_2}^R$ in the ring
\be
\hbox{HL}_{R_1}\{p\}\cdot \hbox{HL}_{R_2}\{p\} = \!\!\!\sum_{{R \in R_1\otimes R_2}\atop{R\in {\tiny \SP}}}\!\!\!
N_{R_1,R_2}^R \hbox{HL}_R\{p\}
\ee
vanish for $R\notin \SP$, provided $t=-1$ and $R_1,R_2\in \SP$.
Note that $\hbox{HL}_R\{p\}$ do not vanish for $R\notin \SP$, and then they can also
depend on even $p_{2k}$, thus the set of $Q_R\{p\}$ is not the same as the {\it set}
of $\hbox{HL}_R$, it is a {\it sub-set}, and a {\it sub-ring}.

There is also a manifest way to construct the $Q$ Schur polynomial as a Pfaffian (see, e.g., \cite[Eq.(74)]{MMN}).

The $Q$ Schur polynomials form a system, which has very close properties to the standard Schur functions, only they form a basis in a {\it sub}space of time-variables. In particular, there is a counterpart of the Frobenius formula (\ref{F}), for the $Q$ Schur polynomials it states
\be\label{FQ}
Q_R\{p_k\}=\sum_{\Delta\in \hbox{\footnotesize OP}}{\Psi_R(\Delta)\over z_\Delta}p_\Delta
\ee
where OP (odd partitions) is a set of Young diagrams with all lengths of lines odd, and $\Psi_R(\Delta)$ are the characters of the Sergeev group \cite{Sergeev,Sergrev}. As any characters, they satisfy the orthogonality conditions:
\be\label{QOR}
\sum_{\Delta\in \hbox{\footnotesize OP}}{\Psi_R(\Delta)\Psi_{R'}(\Delta)\over 2^{l_{_\Delta}}z_\Delta}=\delta_{RR'},
\ \ \ \ \ \ \
\sum_{R\in \hbox{\footnotesize SP}}{\Psi_R(\Delta)\Psi_{R}(\Delta')\over 2^{l_{_\Delta}}z_\Delta}=\delta_{\Delta\Delta'}
\ee
Actually relevant for the $Q$ Schur polynomials is the restriction to odd times,
i.e. the Young diagram $\Delta$ in (\ref{FQ}), which defines the monomial
$p_\Delta = \prod_{i}^{l_\Delta} p_{\Delta_i}$
should have all the lines of odd length:   $\Delta\in \OP$.
Therefore of crucial importance is the celebrated {\bf one-to-one correspondence
between the sets of  $\SP$ and $\OP$}.

At last, the $Q$ Schur polynomials satisfy the Cauchy formula, which we write in a form similar to (\ref{Cauchy}), because this is how it will be used in sec.\ref{Kon} below:
\be\label{QCauchy}
\sum_{R\in {\footnotesize \SP}} Q_{R}\{p\}Q_{R}\{\Tr X^k\}
= \exp\left(\sum_k \frac{p_{2k+1}\Tr X^{2k+1}}{k+1/2}\right)
\ee

\subsection{Combinatorial expression for correlators\label{Cccor}}

We are now ready to formulate a nice result for the correlators (\ref{Kc}) in terms of the $Q$ Schur polynomials. Let us denote through $2R$ the Young diagram produced from the diagram $R$ by doubling all line lengths, and denote through $R|2$ the Young diagram with all even line lengths (in this case, $R/2$ denotes the Young diagram with all line lengths being half of those of $R$)\footnote{For instance, an example of $R|2$ is $R=[6,4,2]$. In this case, $R/2=[3,2,1]$ and $2R=[12,8,4]$.}. Then (compare with \cite{dFIZ}),
\be\label{Qcor}
\boxed{
\left<Q_R\{\Tr X^k\}\right> \ = \ \left\{
\begin{array}{cl}
\displaystyle{{Q_{R/2}\{\Tr\Lambda^{-k}\}Q_{R/2}\{\delta_{k,1}\}\over Q_{R}\{\delta_{k,1}\}}}&\ \ \ \ \ \hbox{if }R|2\cr
&\cr
0&\ \ \ \ \ \hbox{otherwise}
\end{array}
\right.
}
\ee
For the quantity $Q_{R}\{\delta_{k,1}\}$, there is a counterpart of the standard hook formula (\ref{hook}):
\be
Q_{R}\{\delta_{k,1}\}=2^{|R|-{l_{_R}\over 2}}\left({1\over\prod_j^{l_{_R}}R_j!}\right)\prod_{k<m}{R_k-R_m\over R_k+R_m}
\ee
It is non-zero only for $R\in \hbox{SP}$, and
\be
\left<Q_{2R}\{\Tr X^k\}\right> \ = \
\prod_i^{l_R}{(2R_i)!\over 2^{|R|}\cdot R_i!}\cdot Q_{R}\{\Tr\Lambda^{-k}\}
\ee
Note that, unlike most formulas in the Introduction, the r.h.s. in (\ref{Qcor}) is sensitive to
relative normalization of two different $Q$-Schur functions $Q_R$ and $Q_{R/2}$,
which is actually well defined, because these quantities are characters of a group
and they form an algebra with integer Littlewood-Richardson coefficients. Another property that distinguishes this normalization is that,
with the scalar product
\be
\Big< p_{2k+1} \Big| p_{2l+1}\Big> = (k+1/2) \cdot \delta_{k,l}
\ee
the $Q$-polynomials are orthonormal:
\be
\Big< Q_R\Big| Q_{R'} \Big> =\delta_{R,R'}
\ee
and the Cauchy formula (\ref{QCauchy}) acquires an especially simple form with unit coefficients in the sum.

\section{Kontsevich model}

Now we are ready to consider the Kontsevich model given by the integral \cite{Kon}
\be
Z=
\int  \exp\left(- {\Tr X^3\over 3}+\Tr\Lambda^2 X\right) dX
\ee
and it can be further generalized to non-cubic potentials \cite{GKM},
including the quadratic one, when the model is equivalent to the Hermitian matrix model \cite{versus}.
In this paper, we concentrate on the original cubic case, and we denote the background field by $\Lambda^2$
to simplify the formulas in sec.\ref{Kon}.

\subsection{Kontsevich model in the character phase}

The normalization of measure of the Kontsevich integral depends on the phase of the model \cite{GKMU}.
In the character phase, the integral is understood as a formal series in positive powers of $\Tr\Lambda^k$, and the generating function of the correlators is defined to be
\be
Z_{ch}=
{\int dX \exp\left(-{\Tr X^3\over 3}+\Tr\Lambda^2 X\right)\over \int dX \exp\left(-\Tr {X^3\over 3}\right)}
\ee
Since the cubic part does not depend on the angular part of the matrix $X$, one can first perform the angular integration
using the expansion \cite{Mor}
of the Itzykson-Zuber integral \cite{IZ}
\be
\int  \exp\left(\Tr\Lambda^2 UXU^\dagger \right) [dU]=
\sum_R  \frac{\chi_R\{\delta_{k,1}\}}{\chi_R\{N\}}\cdot
\chi_R\{\Tr\Lambda^{2k}\}\chi_R\{\Tr X^k\}
\ee
so that one remains with
\be
Z_{ch}=\sum_R  \frac{\chi_R\{\delta_{k,1}\}}{\chi_R\{N\}}\cdot
\chi_R\{\Tr\Lambda^{2k}\}\left<\chi_R\{\Tr X^k\}\right>
\ee
and then calculate the correlators $\left<\chi_R\{\Tr X^k\}\right>$ for the star-like integration contours directly applying formulas like (\ref{monomodels}).
Thus, in this phase, the expression in terms of characters is straightforward and nearly obvious.

\subsection{Kontsevich model in the Kontsevich phase
\label{Kon}}

Much more interesting is the case of Kontsevich phase, where the character calculus is much less trivial.
In this phase,
the model is given by the integral
\be
Z_K=\exp\left(-{2\over 3}\Tr\Lambda^3\right)
{\int dX \exp\left(- {\Tr X^3\over 3}+\Tr\Lambda^2 X\right)\over \int dX \exp\left(-\Tr X^2\Lambda\right)}
\ee
and is understood as a formal series in $\Tr\Lambda^{-k}$. The simplest way to deal with this integral is to shift the integration variable to the saddle point $X\to X+\Lambda$ so that the integral takes the form
\be\label{ZK}
Z_{K}={\int dX \exp\left(-\Tr {X^3\over 3}-\Tr X^2\Lambda \right)\over \int dX \exp\left(-\Tr X^2\Lambda\right)}
\ee
and one calculates this integral perturbatively using the Cauchy formula (\ref{QCauchy}),
\be
 \exp\left(-{\Tr X^3\over 3}\right)
 = \sum_{R\in \SP} Q_R\{-\Tr X^k\}\cdot Q_{R}\left\{{1\over 2}\delta_{k,3}\right\}
\ee
Since, for even sizes of Young diagrams $|R|$, $Q_R\{-p_k\}=Q_R\{p_k\}$
and averages are non-vanishing only for even sizes, one finally obtains the Kontsevich partition function:
\be\label{KC}
\!\!\boxed{
Z_K=
 \sum_{R\in \SP}{1\over 2^{|R|}} \left<Q_R\{\Tr X^k\}\right>\cdot Q_{R}\left\{\delta_{k,3}\right\} \ =\
\sum_{R\in \SP}{1\over 4^{|R|}}
{Q_{R}\{\Tr\Lambda^{-k}\}Q_{R}\{\delta_{k,1}\}Q_{2R}\{\delta_{k,3}\}\over Q_{2R}\{\delta_{k,1}\}}}
\ee
where we used  (\ref{Qcor}) for the average. Note that, because of the factor $Q_{2R}\{\delta_{k,3}\}$, this sum effectively runs over $R$ of sizes divisible by 3.

As expected in Kontsevich phase, this partition function depends on the matrix size $N$ only through
the variables $\Tr \Lambda^{-k}$.
This is exactly the same phenomenon which we observed in (\ref{compAB}).

Derivation of the formula  (\ref{KC}) could be one of the goals in \cite{dFIZ},
but at that time the $Q$ Schur polynomials were even less known then now, and an explicit answer like
this was unavailable.

\subsection{Integrability and character expansion}

Note that expansion of the partition function in characters typically allows one to check if it is a $\tau$-function of the KP hierarchy immediately by checking that the expansion coefficients satisfy the Pl\"ucker relations. For instance, this is the case for the partition functions (\ref{CP}) and (\ref{GP}). Indeed, any partition function of the form
\be\label{tau2}
Z_w=\sum_RS_R\{p\}S_R\{\bar p\}w_R
\ee
with the function $w_R$ being the product
\be\label{w}
w_R=\prod_{i,j\in R}f(i-j)
\ee
and $\bar p_k$ just arbitrary parameters, is a KP $\tau$-function  \cite{GKM2,OS,AMMN}, since arbitrary Schur polynomial $S_R\{\bar p\}$
satisfies the Pl\"ucker relations, and multiplying any solution to the Pl\"ucker relations by $w_R$ of the form (\ref{w}) preserves the solution. Now using (\ref{toploc}) and fixing $\bar p_k=N_2$, one obtains that (\ref{CP}) is of the form (\ref{tau2}). Similarly, fixing $p_k=\delta_{k,1}$, one obtains that (\ref{GP}) is of the form (\ref{tau2}).

As we explained in \cite[sec.8]{MMN} (see also \cite{Orlov}), similarly looking at expansion in the $Q$ Schur polynomials, one can expect that the partition function is a $\tau$-function of the BKP hierarchy. Moreover, we proposed an example of expansion literally repeating formula (\ref{CP}) but with the Schur polynomials substituted by the $Q$ Schur polynomials, though have not managed to construct a matrix model that possesses such an expansion. Hence, it is surprising to realize that the Kontsevich model, which is a $\tau$-function of the KP hierarchy \cite{GKM}, not BKP have a similar character expansion in the $Q$ Schur polynomials.
This realization in terms of matrix model is especially important as the $Q$ Schur polynomials are related to the spin Hurwitz numbers and to the corresponding cut-and-join operators \cite{MMN}.

Let us emphasize that, as soon as the Kontsevich partition function is a KdV $\tau$-function \cite{GKM}, and the KdV $\tau$-function depends on odd times only, this is quite natural to expand the Kontsevich partition function in the $Q$ Schur polynomials, which, in variance with the ordinary Schur polynomials, depends only on odd times. One may ask why people usually consider expansions of KdV $\tau$-functions into the ordinary Schur polynomials and not into the $Q$ Schur polynomials (see, however, \cite{ON}), which could look more natural. The reason is that the standard technique of the KdV hierarchy as an embedding into the KP hierarchy uses the formalism of free fermions \cite{DJKM}, hence the ordinary Schur polynomials, and it is only using the neutral fermions that gives rise to the $Q$ Schur polynomials, but this type of fermions is associated with the BKP hierarchy \cite{BKP,Orlov}. We will consider this issue in detail elsewhere.

\section{Conclusion}

In this paper, we extended the superintegrability relation $<character> \sim {\rm character}$
to the second kind of matrix models, depending on background fields.
This provides a very nice generalization for the complex matrix model, where dimensions
of matrices are lifted to more general traces of background matrices.
The best known example of the background-field model is, however, different:
this is the Kontsevich model, which is rather analogous to the Hermitian, not complex model.
In the Hermitian case, the simple formulas are known to get a little more involved: they include
characters at a peculiar locus $\delta_{k,s}$, where $s$ is the power of the potential.
Thus, it was an intriguing question what happens in background fields.
Our result in this paper is for the ordinary Kontsevich model with cubic potential,
and it is that the very set of characters is changed from the Schur polynomials to the very interesting set of
$Q$ Schur polynomials, and then their averages are given by a rather transparent formula
(\ref{Qcor}). The formula is, however, somewhat non-trivial, because it requires a correlation
between normalization of different $Q$ Schur polynomials, in most other examples this does not matter
because just a single character appears in the answer which is made from its values
at three different loci.
In the Kontsevich case, this is different, and this adds new colors to the yet-no-so-well-known
subject of $Q$ Schur polynomials.

It is important that the superintegrability-related expansion (\ref{KC}) of the Kontsevich partition function in $Q$ Schur polynomials is much simpler than the expansion through the ordinary Schur polynomials, which could seem natural from the point of view of KP integrability \cite{UFN3,Satsuma}.  Indeed, the coefficients of the latter expansion were found in \cite{Zh,BY} and turned out to be quite complicated and do not expose any clear structure. Even the character nature of the coefficients is not evident in this KP induced expansion. Eq.(\ref{KC}) is obviously free from all such drawbacks, in full accordance with the expectations. It, however, remains a problem to lift (\ref{KC}) to generalized Kontsevich model \cite{GKM}, while the ordinary Schur expansion of them should be a direct lifting of  \cite{Zh,BY}.

\section*{Acknowledgements}

This work was supported by the Russian Science Foundation (Grant No.20-12-00195).


\begin{thebibliography}{12}

\bibitem{MMint} A. Gerasimov, A. Marshakov, A. Mironov, A. Morozov, A. Orlov,
Nucl.Phys. {\bf B357} (1991) 565-618\\
S. Kharchev, A. Marshakov, A. Mironov, A. Orlov, A. Zabrodin,
Nucl.Phys. {\bf B366} (1991) 569-601

\bibitem{versus} S. Kharchev, A. Marshakov, A. Mironov, A. Morozov,
Nucl.Phys. {\bf B397} (1993) 339-378, hep-th/9203043

\bibitem{UFN3} A. Morozov,
Phys.Usp.(UFN) {\bf 37} (1994) 1;
hep-th/9502091; hep-th/0502010\\
A. Mironov, Int.J.Mod.Phys. {\bf A9} (1994) 4355; Phys.Part.Nucl.
{\bf 33} (2002) 537; hep-th/9409190

\bibitem{MM1} A.~Mironov, A.~Morozov,
  Phys.\ Lett.\ {\bf B771} (2017) 503,
arXiv:1705.00976

\bibitem{MM2} A.~Mironov, A.~Morozov,
  Phys.\ Lett.\ {\bf B774} (2017) 210,
arXiv:1706.03667

\bibitem{superint}  A. Mishchenko, A. Fomenko, 
Funct. Anal. Appl. {\bf 12} (1978) 113\\
W. Miller Jr., S. Post, P. Winternitz, 
J. Phys. {\bf A46} (2013) 423001, arXiv:1309.2694

\bibitem{AMMN} A. Alexandrov, A. Mironov, A. Morozov, S. Natanzon,
JHEP 11 (2014) 080,  arXiv:1405.1395

\bibitem{IMM} H.~Itoyama, A.~Mironov, A.~Morozov,
  JHEP {\bf 1706} (2017) 115,
arXiv:1704.08648

\bibitem{KPSS} C. Kristjansen, J. Plefka, G. W. Semenoff, M. Staudacher, Nucl.Phys. {\bf B643} (2002) 3-30, hep-th/0205033

\bibitem{Ramg} S. Corley, A. Jevicki, S. Ramgoolam, Adv.Theor.Math.Phys. {\bf 5} (2002) 809-839, hep-th/0111222

\bibitem{HZ} J. Harer, D. Zagier, Invent.Math. {\bf 85} (1986) 457-485\\
C. Itzykson, J.-B. Zuber, Comm.Math.Phys.
{\bf 134} (1990) 197-208\\
S.K. Lando, A.K. Zvonkin, {\sl Embedded graphs}, Max-Plank-Institut f\"ur Mathematik, Preprint 2001 (63)

\bibitem{HZ1} A.~Morozov and S.~Shakirov,
  JHEP {\bf 0904} (2009) 064,
arXiv:0902.2627

\bibitem{Fulton} W. Fulton, {\sl Young tableaux: with applications to representation theory and geometry},
LMS, 1997

\bibitem{MKR} R. de Mello Koch, S. Ramgoolam, arXiv:1002.1634

\bibitem{MMsum} A.~Mironov, A.~Morozov,
  JHEP {\bf 1808} (2018) 163,
arXiv:1807.02409

\bibitem{BEM} M. Tierz, Mod. Phys. Lett. A19 (2004) 1365-1378, hep-th/0212128\\
A. Brini, B. Eynard, M. Mari\~no, Annales Henri Poincar\'e. Vol. 13. No. 8. SP Birkh\"{a}user Verlag Basel, 2012, arXiv:1105.2012

\bibitem{RJ} M. Rosso, V. F. R. Jones, J. Knot Theory Ramifications, \textbf{2} (1993) 97-112\\
X.-S. Lin, H. Zheng, Trans.Amer.Math.Soc. \textbf{362} (2010) 1-18 math/0601267

\bibitem{AlexMMM} A. Alexandrov, A. Mironov, A. Morozov, An. Morozov,
  JETP Letters {\bf 100} (2014) 271-278 (Pis'ma v ZhETF {\bf 100} (2014) 297-304),  arXiv:1407.3754 \\
A. Alexandrov, D. Melnikov, JETP Letters {\bf 101} (2015) 51–56, arXiv:1411.5698

\bibitem{MPS} A.~Morozov, A.~Popolitov and S.~Shakirov,
  Phys.\ Lett. {\bf B784} (2018) 342,
  arXiv:1803.11401

\bibitem{MMell} A.~Mironov, A.~Morozov,
  arXiv:2011.02855 \\
  A.~Mironov, A.~Morozov,
  arXiv:2011.01762

\bibitem{PSh} C.~Cordova, B.~Heidenreich, A.~Popolitov, S.~Shakirov,
  Commun.\ Math.\ Phys.\  {\bf 361} (2018)   1235,
  arXiv:1611.03142

\bibitem{GKM} S.~Kharchev, A.~Marshakov, A.~Mironov, A.~Morozov, A.~Zabrodin,
  Phys.\ Lett. {\bf B275} (1992) 311,
  hep-th/9111037\\
S.~Kharchev, A.~Marshakov, A.~Mironov, A.~Morozov, A.~Zabrodin,
  Nucl.\ Phys.\ {\bf B380} (1992) 181,
  hep-th/9201013

\bibitem{Kon} M.~Kontsevich,
  Commun.\ Math.\ Phys.\  {\bf 147} (1992) 1

\bibitem{LSZ}
E.~Langmann, R.~J.~Szabo and K.~Zarembo,
JHEP \textbf{01} (2004) 017,
arXiv:hep-th/0308043

\bibitem{IMM3} H.~Itoyama, A.~Mironov and A.~Morozov,
  JHEP {\bf 1912} (2019) 127,
arXiv:1909.06921\\
H.~Itoyama, A.~Mironov and A.~Morozov,
  Phys.\ Lett.\ {\bf B802} (2020) 135237,
arXiv:1910.03261

\bibitem{R} J. Ben Geloun, S. Ramgoolam,
arXiv:1307.6490\\
R. de Mello Koch, D. Gossman, L. Tribelhorn, JHEP, {\bf 2017} (2017) 011, arXiv:1707.01455\\
P. Diaz, S.J. Rey, arXiv:1706.02667\\
J. Ben Geloun, S. Ramgoolam,
arXiv:1708.03524

\bibitem{IMM2} H.~Itoyama, A.~Mironov and A.~Morozov,
  Nucl.\ Phys.\ {\bf B932} (2018) 52,
  arXiv:1710.10027

\bibitem{MMN} A.~Mironov, A.~Morozov, S.~Natanzon,
  Eur.\ Phys.\ J.\ {\bf C80} (2020)  97,
arXiv:1904.11458

\bibitem{Schur} I. Schur,
J. Reine Angew. Math. {\bf 139} (1911) 155-250

\bibitem{Mac}  I.G. Macdonald,
{\it Symmetric functions and Hall polynomials}, Second Edition, Oxford University Press,
1995

\bibitem{Sergeev} A. Sergeev, 
Math. Sb. USSR, {\bf 51} (1985) 419–427

\bibitem{Sergrev} M. Yamaguchi, 
J. Algebra {\bf 222} (1999) 301–327, math/9811090\\
A. Kleshchev, {\sl Linear and projective representations of symmetric groups,} Cambridge
Tracts in Mathematics {\bf 163}, Cambridge Univ. Press (2005)

\bibitem{dFIZ} P.~Di Francesco, C.~Itzykson, J.~B.~Zuber,
  Commun.\ Math.\ Phys.\  {\bf 151} (1993) 193,
  hep-th/9206090

\bibitem{GKMU} A.~Mironov, A.~Morozov, G.~W.~Semenoff,
  Int.\ J.\ Mod.\ Phys.\ {\bf A11} (1996) 5031,
  hep-th/9404005

\bibitem{Mor} A.~Y.~Morozov,
  Theor.\ Math.\ Phys.\  {\bf 162} (2010) 1
   [Teor.\ Mat.\ Fiz.\  {\bf 161} (2010) 3],
arXiv:0906.3518

\bibitem{IZ} Harish-Chandra,
  Am.\ J.\ Math.\  {\bf 79} (1957) 87\\
C.~Itzykson and J.~B.~Zuber,
  J.\ Math.\ Phys.\  {\bf 21} (1980) 411

\bibitem{GKM2} S. Kharchev, A. Marshakov, A. Mironov, A. Morozov, Int.J.Mod.Phys. {\bf A10} (1995) 2015, hep-th/9312210

\bibitem{OS} A. Orlov, D.M. Shcherbin,
Theor.Math.Phys.
{\bf 128} (2001) 906-926\\
A. Orlov,
Theor.Math.Phys. {\bf 146}
(2006) 183–206

\bibitem{Orlov} A. Orlov,  Theor. Math. Phys. {\bf 137} (2003) 1574-1589, math-ph/0302011\\
J. Harnad, J.W. van de Leur, A.Yu. Orlov, Theor. Math. Phys. {\bf 168} (2011) 951-962, arXiv:1101.4216 \\
J.W. van de Leur, A.Yu. Orlov, arXiv:1404.6076 ; arXiv:1611.04577\\
A.Yu. Orlov, T. Shiota, K. Takasaki, arXiv:1201.4518 ; arXiv:1611.02244

\bibitem{ON} J.J.C. Nimmo, A. Orlov, Glasgow Mathematical Journal {\bf 47 (A)} (2005) 149-168, nlin/0405009

\bibitem{DJKM} E. Date, M. Jimbo, M. Kashiwara, T. Miwa, 
RIMS Symp. {\sl "Non-linear integrable
systems - classical theory and quantum theory"} (World Scientific,
Singapore, 1983)

\bibitem{BKP} E. Date, M. Jimbo, M. Kashiwara, T. Miwa,
Physica {\bf D4} (1982) 343-365\\
M. Jimbo, T. Miwa,
Publ. RIMS Kyoto Univ. {\bf 19} (1983) 943-1001

\bibitem{Satsuma} Y.Ohta, J.Satsuma, D.Takahashi, T.Tokihiro,
Prog.Theor.Phys.Suppl. {\bf 94} (1988) 210

\bibitem{Zh} J.~Zhou,
  arXiv:1306.5429

\bibitem{BY} F.~Balogh and D.~Yang,
  Lett.\ Math.\ Phys.\  {\bf 107} (2017) 1837, arXiv:1412.4419

\end{thebibliography}
\end{document}